\documentclass[proof]{WileyASNA-v1}

\newcommand{\logg}{\mbox{$\log g$}}
\newcommand{\Teff}{\mbox{$T_\mathrm{eff}$}}
\newcommand{\Msol}{$M_\odot$}

\articletype{Article Type}%

\received{}
\revised{}
\accepted{}

\begin{document}

\title{Searching the non-accreting white dwarf population in eROSITA data}

\author[1]{S. Friedrich*}

\author[1]{C. Maitra}

\author[1]{K. Dennerl}

\author[2]{A. Schwope}

\author[3]{K. Werner}

\author[3]{B. Stelzer}

\authormark{S. Friedrich \textsc{et al}}

\address[1]{\orgname{Max-Planck-Institut f. extraterrestrische Physik}, \orgaddress{\state{Garching}, \country{Germany}}}

\address[2]{\orgname{Leibniz-Institut für Astrophysik Potsdam}, \orgaddress{\state{Potsdam}, \country{Germany}}}

\address[3]{\orgdiv{Institut f\"ur Astronomie \& Astrophysik}, \orgname{Eberhard-Karls-Universit\"at T\"ubingen}, \orgaddress{\state{T\"ubingen}, \country{Germany}}}

\corres{*S. Friedrich, Max-Planck-Institut f. extraterrestrische Physik, Garching, Germany. \email{sfriedrich@mpe.mpg.de}}

\presentaddress{Giessenbachstr. 1, D-85748 Garching, Germany}

\abstract{eROSITA is the soft X-ray instrument aboard the Spectrum Roentgen Gamma (SRG) satellite that is most sensitive in the energy range between 0.2 and 2.3 keV. Between December 2019 and December 2021 eROSITA completed four all-sky surveys producing all-sky X-ray source lists and sky maps of unprecedented depth. In the energy range between 0.2 keV and 1 keV, we detected about 38,000 sources with a hardness ratio below $-0.94$, covering a small sample of known white dwarfs found with eROSITA in the dataset to which the German eROSITA consortium has rights (half sky). 264 of these soft sources have a probability of more than 90 \% to be a white dwarf. This is more than the 175 white dwarfs ROSAT found in the whole sky. Here we present the results of a pilot study to increase the sensitivity of eROSITA for soft sources by extending the detection threshold down to 0.1 keV. First tests with dedicated sky regions are promising.}




\maketitle


\section{Introduction}\label{sec1}
SRG/eROSITA (extended ROentgen Survey with an Imaging Telescope Array, \cite{Predehl2021}) performed the first all-sky X-ray survey almost 30 years after the legendary soft X-ray survey by ROSAT \citep{Truemper1982}. Between December 2019 and December 2021 four all-sky surveys were completed. Although eROSITA is designed to detect very large samples ($\approx$100,000 objects) of galaxy clusters up to redshifts z $\ge$ 1 as extended X-ray sources, it detects at the same time an unprecedentedly large number of unresolved, point-like X-ray sources. The catalogues compiled by the German eROSITA consortium are used to generate large samples of all types of X-ray emitters, from stellar coronal emitters to galactic compact binaries to AGN at large cosmological distances \citep{Merloni2024, freund+24, schwope+24, salvato+22, wolf+23}. Compared to ROSAT, eROSITA has a higher sensitivity in the soft band below 2 keV. On the other hand, the sensitivity of eROSITA at very soft energies of about 0.1 keV is not as good as ROSAT’s. However, the larger effective area (1237 cm$^2$ at 1 keV) and the typically longer exposure (four surveys instead of only one) compensate for this. Unfortunately, the two of the seven cameras of eROSITA which were designed to be sensitive to energies lower than 0.2~keV suffer from optical stray light \citep{Predehl2021}, and their use for detecting soft sources needs more detailed investigations. Until this is achieved we restrict our tests to the "non-light leak" telescope modules TM 1$-$4, and TM 6.

Single (i.e., non-accreting) white dwarfs belong to one of the rarer classes of objects detected in X-ray surveys. ROSAT detected only 175 white dwarfs \citep{Fleming1996}, a number far below the previous expectations. The majority (161) are hydrogen-rich (DA) white dwarfs with effective temperatures in excess of about \Teff\ = 30,000\,K. At these temperatures, hydrogen in the atmospheres is sufficiently ionized and its opacity is so low that thermal radiation from deep hot layers escapes the star. Most DAs with temperatures below 40,000\,K contain no material in their atmospheres other than hydrogen. However, at higher temperatures, namely beyond about 50,000--60,000\,K, trace metals are sustained in the atmosphere by radiative levitation, and their opacities effectively block the soft-X-ray flux \citep[e.g.,][]{Barstow+1993}.

ROSAT also detected a small number of non-DA (i.e., hydrogen-deficient, helium-dominated) white dwarfs (spectral type DO) and their immediate progenitors, the PG1159 stars. The DO white dwarfs must be significantly hotter than the DAs to emit soft X-rays, because otherwise, the He\,II ground state edge blocks the flux at wavelengths below 228\,\AA. Only at temperatures above about 100,000\,K, helium is sufficiently ionised to He\,III and the atmosphere becomes transparent for X-rays. Thus, only two DOs were detected by ROSAT and their temperatures are \Teff = 115,000\,K \citep[PG1034+001,][]{Werner+2017} and \Teff = 200,000\,K \citep[KPD0005+5106,][]{Werner+2015}. PG1159 stars, the DO progenitors, have high amounts of carbon and oxygen in their atmospheres and their opacities become low only at very high temperatures. ROSAT discovered eight PG1159 stars, all with temperatures of 140,000\,K and higher. One example is WD1144+004 which is discussed in Sect.\,\ref{sect:pg1144}.

\section{White dwarfs in the current eROSITA processing}\label{sec2}

\begin{figure}
	\centerline{\includegraphics[width=90mm,height=90mm]{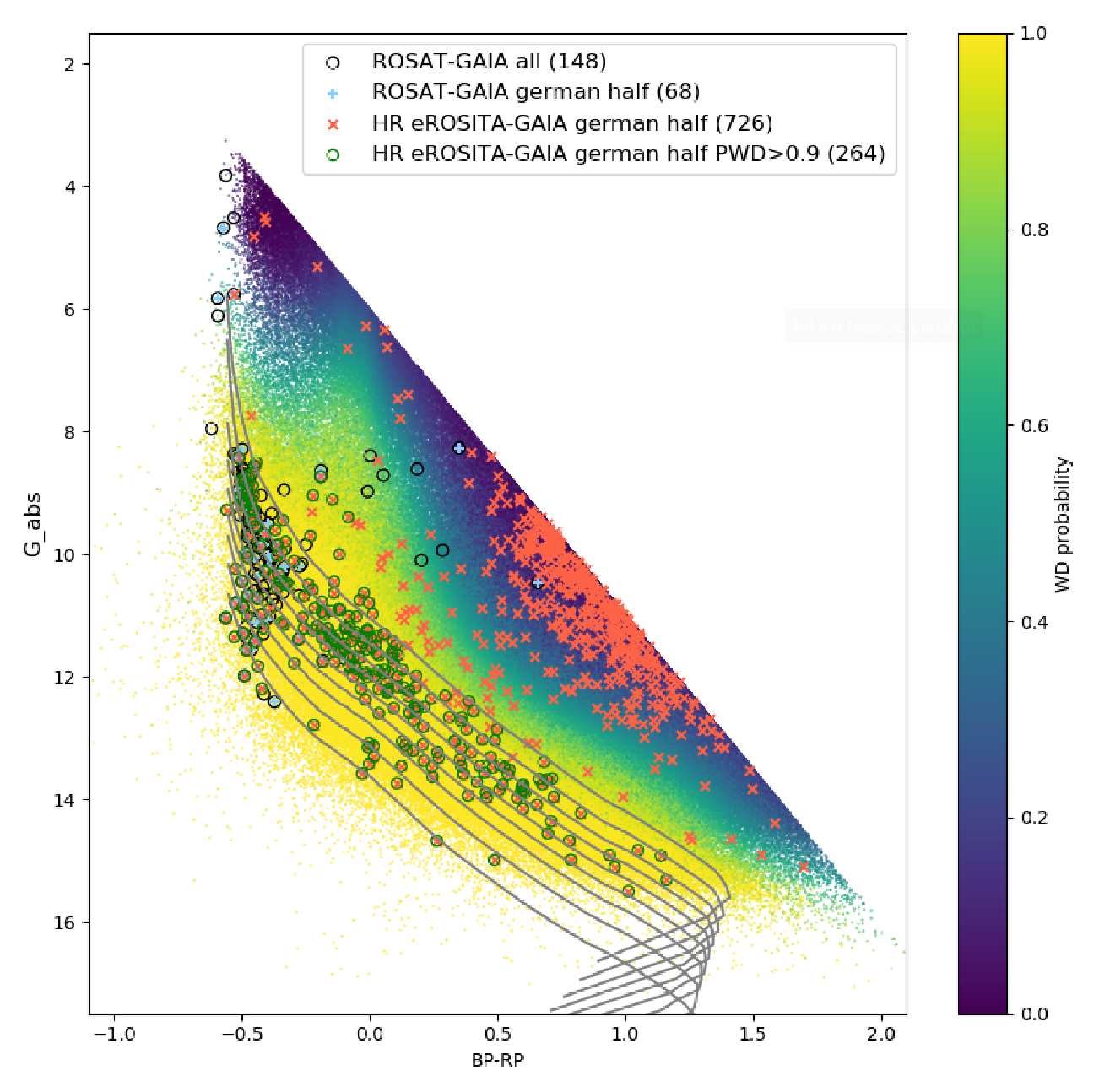}}
	\caption{All white dwarfs from the {\it Gaia} white dwarf catalogue of \cite{Gentile2021} colour-coded with the probability of being a white dwarf. Black circles and light blue crosses mark sources in the  ROSAT white dwarf catalogue \citep{Fleming1996} that matches the {\it Gaia} white dwarf catalogue and which are in the German half of the sky, respectively. Orange crosses mark eROSITA sources with a hardness ratio of $\le -0.94$ and with counterparts in the {\it Gaia} white dwarf catalogue. Those highlighted with a green circle have a probability larger than 90\,\% of being a white dwarf in that catalogue. Overplotted in grey are cooling tracks of white dwarfs with masses between 0.4 and 1.2 solar masses in steps of 0.1 solar masses (top to bottom) from the La Plata Stellar Evolution  and Pulsation Research Group \citep{Althaus2013,Camisassa2019,Camisassa2016}. The model atmospheres are from \cite{Koester2010}.\label{fig1}}
\end{figure}

We took the current collaboration-internal preliminary cumulative eROSITA catalogue from the surveys 1$-$4 (German half only), which was processed with the eROSITA Science Analysis Software System (eSASS, \cite{brunner2022}, \texttt{eSASSusers\_240410}). For compiling this catalogue only photons above 0.2~keV were used. From this catalogue, we then selected point sources (extension likelihood = 0) with detection likelihood in the soft band $>3$ and hardness ratios
\begin{eqnarray}
    \frac{cts_{\left[ 0.5 \mathrm{keV} - 1.0 \mathrm{keV}\right]} - cts_{\left[ 0.2 \mathrm{keV} - 0.5 \mathrm{keV} \right]}}{cts_{\left[ 0.5 \mathrm{keV} - 1.0 \mathrm{keV} \right]} + cts_{\left[ 0.2 \mathrm{keV} - 0.5 \mathrm{keV} \right]}} \le - 0.94
\end{eqnarray}

This threshold was determined using a small sample of 53 white dwarfs that have a probability of 1 being a white dwarf in the {\it Gaia} white dwarf catalogue \citep{Gentile2021} and were also discovered with eROSITA. A bimodal distribution was found with hardness ratios either less than $-0.94$ or greater than $-0.34$. Since single white dwarfs should have hardness ratios of about $-1$, we chose $-0.94$ as our threshold. 38,080 soft sources were detected with hardness ratios below $-0.94$. From this sample, we found 726 sources that have matches with the {\it Gaia} white dwarf catalogue \citep{Gentile2021} using NWAY \citep{Salvato2018}. 264 of them have a probability of higher than 90 \% being a white dwarf in this catalogue. This is more than the 175 white dwarfs ROSAT detected in the whole sky \citep{Fleming1996}.

In Figure~\ref{fig1} we show the colour-magnitude diagram of white dwarf candidates detected with {\it Gaia} \citep{Gentile2021} colour-coded with their probability of being a white dwarf. Overplotted are 148 white dwarfs from the ROSAT white dwarf catalogue (black circles) that have counterparts in the {\it Gaia} white dwarf catalogue \citep{Gentile2021} of which 68 are in the German half of the sky (light blue crosses). Orange crosses depict eROSITA sources with a hardness ratio of $\le -0.94$ and counterparts in the {\it Gaia} white dwarf catalogue \citep{Gentile2021}. Those highlighted with a green circle have a probability larger than 90\,\% of being a white dwarf in that work. The diagram clearly shows that a significant fraction of the 38,080 soft sources have a probability of being a WD of less than 40\,\%. As yet we did not perform any studies or observations to identify these objects. Possible candidates are cataclysmic variables and white dwarfs in binary systems. It is also clear that in the yellow area where the WD probability approaches 1, eROSITA detects objects almost 3 magnitudes fainter than ROSAT.

\begin{figure*}[t]
    \centerline{\includegraphics[width=160mm,height=106mm]{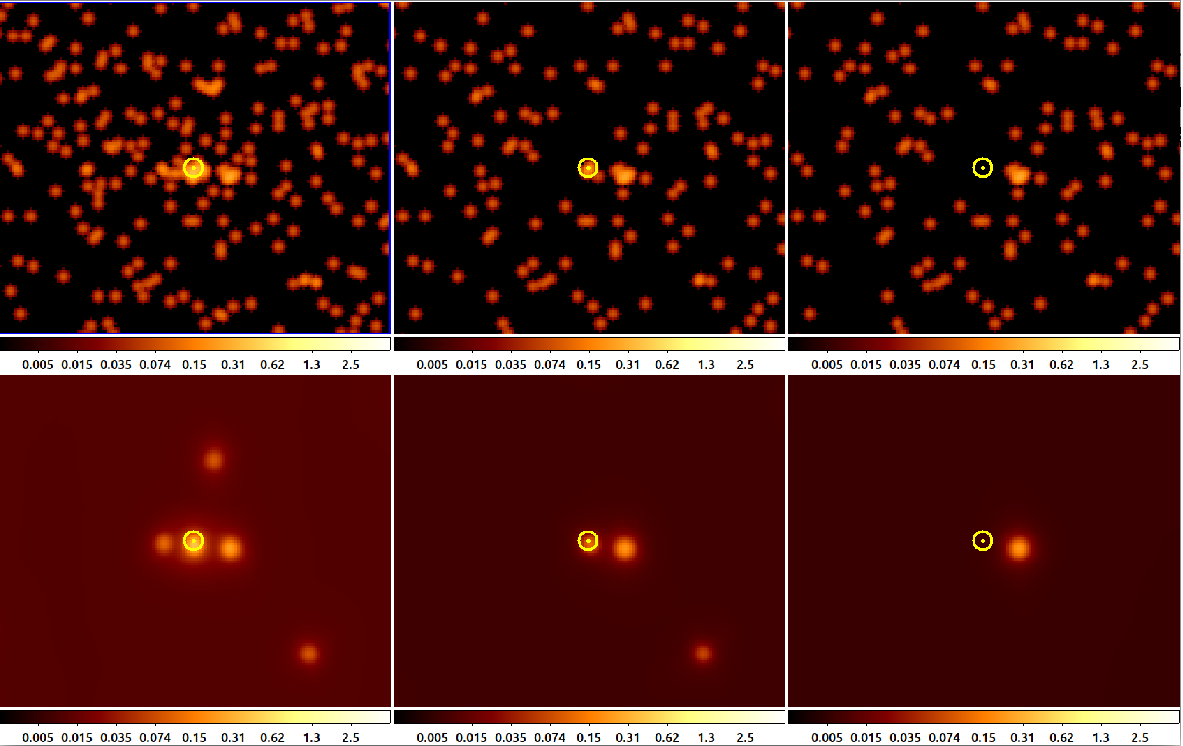}}
    \caption{Sky region with WD0631+107 in three different energy bands from left to right 
    $0.1 - 2.3$, $0.125 - 2.3$, $0.2 - 2.3$\,keV.
    The upper panel shows event images, and the lower one the source model.  The yellow dot and circle mark the source position and an error circle of 30 arcsec. WD0631+107 is not detected in the $0.2-2.3$\,keV range.\label{fig2}}
\end{figure*}

\section{Special processing: 0.1--2.3 keV}\label{sec3}
For the generation of published X-ray catalogs \citep{brunner2022, Merloni2024} only photons in the energy band 0.2$-$2.3 keV were used. This band was chosen to optimize completeness and minimize spuriosity of the bulk of the sources, stellar coronal emitters, and AGN. It is not optimized to detect white dwarfs (or other soft sources). To increase the sensitivity for white dwarfs and other soft sources we ran tests on selected sky regions (so-called sky tiles) using TM 1$-$4, and 6, with a lower energy limit of $E_{\rm min} = 0.1$ and $0.125$ keV (hereafter called special processing). Otherwise, we used the same parameter settings as in the standard eSASS \citep{brunner2022}. 

While visual inspection of the image with $E_{\rm min} = 0.125$\,keV revealed a similar background level as in the image with $E_{\rm min} = 0.2$\,keV, the image with $E_{\rm min} = 0.1$\,keV showed a significantly increased background below the detection threshold. We, therefore, plan to perform simulations with SIXTE \citep{Dauser2019} to optimize the source detection efficiency for soft sources given the increased background at low energies and thus find the optimum value of $E_{\rm min}$.

\section{Test sample selection}\label{sec4}
To find good test fields we compared the latest cumulative eROSITA catalogue from surveys 1$-$4 (called eRASS:4) with known white dwarf catalogues from ROSAT \citep{Fleming1996}, \cite{McCook1999}, and {\it Gaia} \cite{Gentile2021}. Eventually, we have chosen five sky tiles \citep{Predehl2021,Merloni2024} that have an identical entry in all of the named catalogues.
In one of these, the (only) WD in the field was also detected with the standard eSASS.

\subsection{WD0346--011 (GD 50)}
This DA white dwarf was detected by the standard eSASS ($>$0.2 keV) and is listed in all catalogues mentioned above. From its exceptionally high surface gravity, it was found to be a very massive white dwarf \citep{Bergeron+1991}. A recent analysis of the Balmer line spectrum arrived at \Teff\ = 37,535\,K, \logg\ = 9.1, and $M=1.25$\,\Msol \citep{Jewett+2024}.

\subsection{WD1144+004 (PG1144+005)}\label{sect:pg1144}
This (hydrogen-deficient) PG1159 star is not detected by the source detection algorithm of the standard eSASS nor by the special processing. A nearby soft source which is only seen in the special processing may cause some source confusion. The star is very hot (\Teff\ = 150,000\,K) and its atmosphere is mainly composed of carbon and helium \citep{Werner+2016}.

\subsection{WD0631+107 (KPD0631+1043)}
This DA white dwarf is not detected in the standard eSASS but with the source detection algorithm of the special processing (Figure~\ref{fig2}). With \Teff\ = 27,630\,K \citep{Gianninas+2011}, the star is among the coolest white dwarfs with detected soft X-ray emission \citep{Kidder+1992}.

\subsection{WD1125--025 (PG1125--025)}
This DA white dwarf is not detected in the standard eSASS but with the source detection algorithm of the special processing. Its effective temperature was determined to \Teff\ = 31,755\,K \citep{koester+2009}.

\subsection{WD2020--425}
The DA white dwarf is detected by the standard eSASS and detected by the source detection algorithm of the special processing. \cite{koester+2009} measured a temperature of \Teff\ = 28,412\,K.

\section{Conclusions}\label{sec3}
We could show, that significant detections of soft sources can be made with eROSITA by lowering the energy threshold for source detection. Experiments were made with special processing down to 0.1\,keV to find soft sources not detected by the standard processing. Comprehensive tests followed by systematic processing of all-sky tiles with German data rights will be used to compile a flux-limited sample of isolated white dwarfs. This new catalog of soft sources will contain other notorious soft X-ray emitters like isolated neutron stars \citep{kurpas+24}, accreting white dwarfs \citep[aka CVs, polars and soft IPs][]{schwope+24}, Super Soft Sources \citep{maitra+24}, stellar coronae \citep{freund+24} and supersoft AGN \citep{boller+96}. In addition, the analysis of white dwarfs such as GD\,153 and HZ\,43 in this way will facilitate cross-calibration with other X-ray telescopes, e.g. {\it Chandra} or ROSAT.


\section*{Acknowledgments}

This work is based on data from eROSITA, the soft X-ray instrument aboard SRG, a joint Russian-German science mission supported by the Russian Space Agency (Roskosmos), in the interests of the Russian Academy of Sciences represented by its Space Research Institute (IKI), and the Deutsches Zentrum für Luft- und Raumfahrt (DLR). The SRG spacecraft was built by Lavochkin Association (NPOL) and its subcontractors, and is operated by NPOL with support from the Max Planck Institute for Extraterrestrial Physics (MPE).
The development and construction of the eROSITA X-ray instrument was led by MPE, with contributions from the Dr. Karl Remeis Observatory Bamberg \& ECAP (FAU Erlangen-Nuernberg), the University of Hamburg Observatory, the Leibniz Institute for Astrophysics Potsdam (AIP), and the Institute for Astronomy and Astrophysics of the University of Tübingen, with the support of DLR and the Max Planck Society. The Argelander Institute for Astronomy of the University of Bonn and the Ludwig Maximilians Universität Munich also participated in the science preparation for eROSITA.
The eROSITA data shown here were processed using the eSASS/NRTA software system developed by the German eROSITA consortium.

\subsection*{Author contributions}

All authors discussed the results and contributed to the final manuscript.

\subsection*{Financial disclosure}

None reported.

\subsection*{Conflict of interest}

The authors declare no potential conflict of interest.

\bibliography{eroWD}%

\end{document}